# Detecting Threat E-mails using Bayesian Approach

M. Tariq Banday[1], Jameel A. Qadri[2], Tariq. R. Jan[3], Nisar. A. Shah[4]

[1] P. G. Department of Electronics and Instrumentation Technology,
University of Kashmir, Srinagar - 6, India
Email: sgrmtb@kashmiruniversity.ac.in

[2] BC College of North West London,
470 Church Lane, Kingsbury, NW9 8UA, London
Email: scorpiojameel@yahoo.com

[3] P.G. Departments of Statistics,
University of Kashmir, Srinagar - 6, India,
Email: trj527@gmail.com

[4] P. G. Department of Electronics and Instrumentation Technology,
University of Kashmir, Srinagar - 6, India
Email: nassgr@yahoo.com

**Abstract:** *Fraud and terrorism have a close connect in terms of the processes that enables and promote them. In the era of Internet, its various services that include Web, e-mail, social networks, blogs, instant messaging, chats, etc. are used in terrorism not only for communication but also for i) creation of ideology, ii) resource gathering, iii) recruitment, indoctrination and training, iv) creation of terror network, and v) information gathering. A major challenge for law enforcement and intelligence agencies is efficient and accurate gathering of relevant and growing volume of crime data. This paper reports on use of established Naïve Bayesian filter for classification of threat e-mails. Efficiency in filtering threat e-mail by use of three different Naïve Bayesian filter approaches i.e. single keywords, weighted multiple keywords and weighted multiple keywords with keyword context matching are evaluated on a threat e-mail corpus created by extracting data from sources that are very close to terrorism.*

**Keywords:** *SPAM, Threat e-mail, Bayesian Filters, E-mail Classification*

## 1. Introduction

Terrorism, which is politically and emotionally charged, involves violence and threat of violence [1, 2]. The most lethal terrorist attack in history commonly known as 9/11 which tool lives of 3000 Americans and international citizens led to far reaching changes in anti-terror approaches and operations in the US and around the Globe [3]. Several law enforcement and intelligence agencies are actively collecting and analyzing crime data to detect and prevent future attacks. Various resource centers and databases like TPDRC [4] and GTD [5] have been created to archive and distribute data collected by government agencies, non-government agencies and researchers about terrorism to extent research and administrative data from across the world that are relevant to the study of terrorism.

E-mail has emerged as a free, valuable and crucial worldwide business tool that not only supports conversation between parties but also supports delivery of documents and archives of diverse nature. Over the last few years, the arteries of e-mail have become literally clogged by spam, viruses, and any other content that can be sent via e-mail. Spam e-mails often contain offensive, fraudulent, adult oriented and misleading material that cause several problems [6] either directly or indirectly to the e-mail system that include: i) network conjunction, ii) misuse of storage space and computational resources, iii) loss of work productivity and annoyance to users, iv) legal issues as a result of pornographic advertisements and other objectionable material, v) financial losses through phishing and other related attacks, vi) spread of viruses, worms and Trojan Horses, and vii) Denial of Services and Directory Harvesting attacks. Several anti-spam procedures have been proposed that try to tackle the problem of spam at various levels in the system [7]. These procedures propose the use of diverse technological, legal, social and economical solutions. A high level review of spam filtering procedures is provided in [8, 9].

E-mail is also misused to send threat e-mails and disseminate other objectionable material related to terrorism. Threat e-mails send threat of an attack to people or government in order to create panic or disturbance. Further, e-mails that disseminate material objectionable to governments like terrorism propaganda for creation of ideology, resource gathering and fund raising may also be classified as threat e-mails. Threat e-mails may be classified by spam filters as legitimate or spam depending upon their content. However, for law enforcement and intelligence agencies a threat e-mail is a source of investigations regardless of the threat being true or false and thus cannot be treated as spam. Anti-terror approaches can be strengthened by monitoring e-mail archives of suspects and installations of 'threat e-mail detectors' at mail servers. Several studies have been carried out to compare and contrast relative efficiency of learning based filters; however, limited literature is available on use of filters for detection of threat e-mail. S. Appavu et al in their research works [10-13] have used Decision Tree, SVM, NN and Naïve Bayes approaches for detecting threat e-mail. These studies compared efficiency and effectiveness of some filters in classifying threat e-mails and reported that Decision Tree based filters outperformed others. A limitation with these studies non-availability of threat e-mail corpus.

In this study instead of using self created threat e-mail messages, we used data that was extracted from various sources which either belongs to terrorists or keep record of terrorism or report terrorism or have some form of interaction with them. We used terrorism websites, newspapers, data available with resource centers and databases like TPDRC and GTD to create an e-mail corpus and applied Naïve Bayes filtering approaches for its classification. The remaining paper is organized as follows: Section 2 introduces e-mail filter, its operation and Bayesian filtering approach. Section 3 details on the corpus and Bayesian approaches used in our experiments. It presents results obtained through experiments on the threat e-mail corpus and analyzes them in terms of various metrics used to evaluate them. We draw conclusion and present future research motivation in section 4.

## 2. E-mail Filters

An e-mail message consists of two parts, namely header and body. Header is a structured set of fields, each having a name and specific meaning. It includes fields namely From, To, Subject, CC, BCC, etc. Message body generally referred to as content of the message is usually text, possibly with HTML markup and MIME encoded attachments. Message analysis and filtering involves selection of features from header and/or body or from message as a whole relevant for analysis. A filter may check the presence of certain words or may consider the arrival of a dozen of substantially identical messages in a certain slot of time. In addition to this, a learning-based filter analyzes a collection of labeled training data which are pre-collected messages with reliable judgment.

A spam filter in general is an application that implements a function to classify an incoming e-mail message as spam or legitimate mail using a particular classification method. Such a system implements the following function:

$$f(m, \theta) = \begin{cases} C_{spam}, & \text{if classified as SPAM} \\ C_{leg}, & \text{otherwise} \end{cases}$$

*Where, $m$ is the message to be classified, $\theta$ is a vector of parameters, and $C_{spam}$ and $C_{leg}$ are respectively spam and legitimate messages.*

Most of Spam filters including the statistical spam filters use machine learning classification techniques wherein the vector of parameter $\theta$ is the result of training the classifier on a pre-collected dataset which may be rebuilding itself with every new message. $\theta$ for such filters can be defined as:

$$\theta = \Theta(X),$$

Learning-based spam filters treat the input data as an unstructured set of tokens, filtering can be applied either to the whole message or to any part of it. For this group of filters with two classes of messages: spam and legitimate mail, there exists a set of labeled training messages, each message being a vector of $d$ binary features and each label being $C_{spam}$ or $C_{leg}$ depending on the class of the message. The training data set **M**, once pre-processed in this way, can be described as:

$$X = \{(\bar{x}_1, y_1), (\bar{x}_2, y_1), \dots (\bar{x}_n, y_n)\}$$

$$\bar{x}_i \in \mathbb{Z}_2^d, \qquad y_i \in \{C_{spam}, C_{leg}\}$$

*Where, $d$ is the number of features used. $\bar{x}_i \in \mathbb{Z}_2^d$ is a new sample the classifier should provide a decision $y \in \{C_{spam}, C_{leg}\}$. $y_1, y_2, \dots y_n$ and labels and $\Theta$ being the training function.*

### 2.1. Filter Model

Learning based e-mail filter consist of learning and detection stages as shown in figure 1. The learning stage uses a training set in the form of known spam and legitimate e-mails collectively called e-mail corpus. Features are extracted from each e-mail of corpus, which are then reduced by a feature reduction function. A training function calculates likelihood probability of each feature occurring in spam and legitimate e-mails. It also calculates the prior probability of each class i.e. spam and legitimate. The feature sets along with their likelihood probabilities are stored in a library for use in the detection stage. A new e-mail message is parsed in the detection stage with respect to the features in the feature set and group probabilities of each group are calculated. If the total calculated probability of spam is greater than some predefined threshold value, the mail is classified as spam otherwise it is classified as legitimate. The feature set and feature probability library is updated with every new classified e-mail.

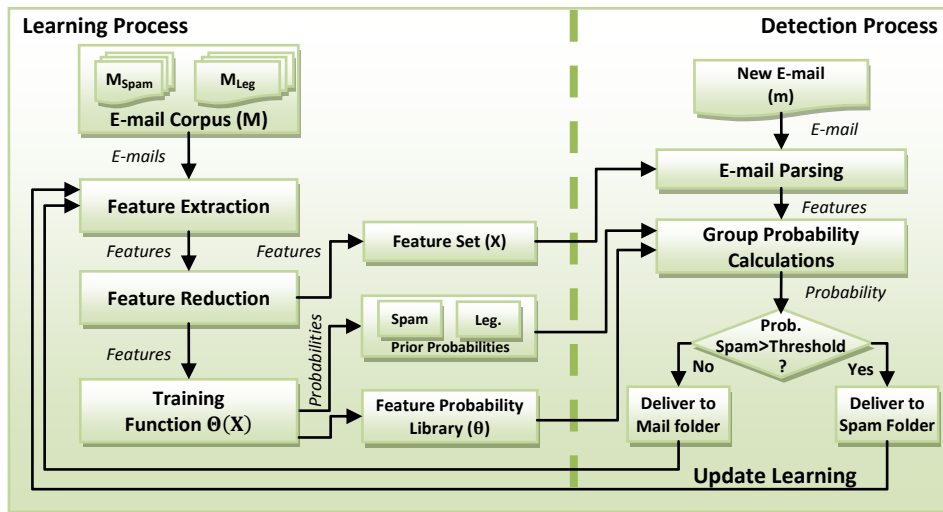

**Figure** 1: Learning E-mail Filter Model

## 2.2. Bayesian Filtering Approach

Bayes theorem also known as Bayes reasoning is used to solve problems which involve conditional probabilities. It was published by Thomos Bayes (1702-1761) and calculates posterior probability based on the probabilities of previous samples. The theorem has been used to classify an e-mail into spam and ham (legitimate) classes by many researchers using different approaches. Most prominent works are of Sahami and Androutsopoulos [14], Paul Graham's [15] and Gary Robinson [16]. Using Bayesian filtering, an e-mail message $C$ presented by a set of $k_x$ features $F_x = \{f_1, f_2, f_3 \ldots f_{k_x}\}$ can be classified into a particular class $c_k$ (e.g. spam ($c_{spam}$) and legitimate ($c_{leg}$)) as follows:

$$P(C = c_k | F = F_x) = \frac{P(F = F_x | C = c_k) P(C = c_k)}{P(F = F_x)}$$

Where $P(F = F_x | C = c_k)$ is the likelihood Probability of $F_x$ occurring in $c_k$. $P(C = c_k)$ represents a prior probability of an e-mail being in class $c_k$ calculated from training examples. $P(F = F_x)$ is the probability of $F_x$ occurring in training examples which will remain constant for an e-mail message while analyzing it.

It is hard to estimate the likelihood probability of $P(F = F_x | C = c_k)$ because it needs complete coverage of the feature space in order to calculate probabilities. Under Paul Graham's [15] Naïve Bayesian assumption that attribute values in feature set $F_x$ are conditionally independent given the target value of spam or ham (legitimate); this probability is calculated as the product of probabilities of the individual attributes. Thus $P(C = c_k | F = F_x)$

$$= \frac{\prod_i P(F_i = f_i | C = c_k) P(C = c_k)}{\prod_i P(F_i = f_i)}$$

$$= P(C = c_k) \prod_i \left( \frac{P(F_i = f_i | C = c_k)}{P(F_i = f_i)} \right)$$

Prior probability $P(C = c_k)$ can be estimated from the training set as:

$$P(C = c_k) = \frac{Number\ of\ e-mails\ belonging\ to\ class\ c_k}{Total\ number\ of\ e-mails\ in\ the\ training\ corpus}$$

If $c_k = \{c_{spam}, c_{leg}\}$, $ngood$ is the number of legitimate e-mails and $nbad$ is the number of spam e-mails in the training corpus then prior probability $P(C = c_{spam}) = \frac{nbad}{nbad+ngood}$.

The training set can also be used to estimate conditional probability of individual attributes $P(F_i = f_i|C = c_k)$ as follows:

$$P(F_i = f_i|C = c_k) =$$

$$\frac{Frequency\ count\ of\ feature\ f_i\ appears\ in\ the\ class\ c_k}{Number\ of\ e-mails\ belonging\ to\ class\ c_k}$$

If $hb(f_i)$ spam e-mail and $hg(f_i)$ legitimate e-mails contain a feature $f_i$ then for an e-mail randomly selected from $M$, a simple estimation for $P(F_i = f_i|C = c_{Spam})$ can be obtained as $\frac{hb(f_i)}{nbad}$. Further, by definition the probability of $f_i$ occurring in training examples $P(F_i = f_i)$ is given by $\frac{hb(f_i)+ hg(f_i)}{ngood+nbad}$. Paul Graham's [15] calculates $P(F_i = f_i|C = c_{Spam})$ as follows:

$$P(F_i = f_i|C = c_{Spam}) = P(f_i) = \frac{b(f_i)}{b(f_i) + g(f_i)}$$

Where $b(f_i) = \frac{hb(f_i)}{nbad}$ and $g(f_i) = \frac{hg(f_i)}{ngood}$.

## 3. Experimental Setup and Analysis

### 3.1. Testing Corpus

To our knowledge no dataset of threat e-mails is publically available in form of a corpus for analysis. Researchers are thus forced to create their own threat e-mail dataset which makes the exact analysis of filtering algorithms difficult. In order to make our analysis more accurate, we choose to create threat e-mail corpus from sources that are close to terrorism. We selected e-mail corpus containing 7000 e-mail messages of which 2700 were legitimate, 2700 were spam and 1600 were threat with no messages having its duplicates in the dataset. Threat e-mail were created from multiple sources that include: data available on terrorist websites, names of terrorist organizations, names of top terrorists, data available in newspapers pertaining to terrorist attacks, threats to government and appeals for shutdowns received by a newspaper for publications, etc. Maximum size of the message was limited to 200 words. Legitimate and spam e-mails received by the authors in their e-mail accounts along with messages created from data present in the websites and news pertaining to terrorism other than threat messages were used as a source for non-threat e-mails. We also used publically available e-mail filtering corpus LingSpam as a source of non-threat e-mails for evaluation. It produced results close to those obtained with the reported dataset. We, in our experiment used different data sizes with different number of features. 75% of the messages from each set were used for training and the remaining 25% messages were used for evaluation. Spam and legitimate messages were collectively used as non-threat messages. A few sample threat e-mails used in our experiments are given in appendix.

### 3.2. Feature extraction and reduction

The subject and the body fields of each e-mail message in the corpus were tokenized based on the space and punctuation. Further, we removed tokens of less than 4 characters which make text classification easy by eliminating words that are found frequently in a list. These stop list words include words such as a, as, the, for, on, etc. which are not useful in classification. We used Porter Steaming algorithm for word steaming to convert words to their morphological base automatically. Stopword removal and word steaming considerably reduces feature space and improves prediction accuracy. One keyword, two keyword and three keyword tokens are extracted and used as features using Term Frequency (TF) vector. Top most valuable features extracted in feature extraction are selected on the basis of information gain (IG) of each feature. Next k-mean algorithm is used to partition the feature space into four groups of similar features.

### 3.3. Bayesian approaches used in Experiments

A feature selected as a token may be a single keyword with or without weights assigned to them or combination of two or more keywords with different weights assigned to multi-keywords. Further, as proposed in [17] keyword contexts can be added to the weighted spam score to make the classification more accurate. Accordingly three different approaches namely single keyword, weighted multiple keyword and weighted multiple keyword with keyword context matching are possible.

In the first approach, weights are either not applied to keywords or same weights are applied to single and multi-keywords. The Bayesian threat score $T_{score}$ is calculated by adding the individual keyword score $T(F_i = f_i)$ for all keywords. This can be mathematically expressed as:

$$T_{score} = \sum_{i=1}^{k_x} T(F_i = f_i)$$

In the second approach, different weights are assigned to tokens having different number of keywords in it e.g. a weight of $W \times 2$ to tokens of two keywords and $W \times 3$ for tokens of 3 keywords. Tokens containing more keywords in it are assigned higher weights than those having fewer keywords. Weights assigned to tokens $W((F_{ij} = f_{ij})$ multiplied by individual keyword scores $T(F_{ij} = f_{ij})$ are summed up to get the Bayesian weighted threat score $T_{WScore}$ as shown below:

$$T_{WScore} = \sum_{i=1, j=1}^{i=k_x, j=n} T(F_{ij} = f_{ij}) \times W((F_{ij} = f_{ij})$$

In the third approach, keyword context score is added to the weighted Bayesian score for all keywords found in the e-mail. Thus Bayesian context weighted threat score $T_{CWScore}$ is thus calculated as:

$$T_{CWScore} = \sum_{i=1,j=1}^{i=k_x,j=n} W_1(T(F_{ij} = f_{ij}) \times W((F_{ij} = f_{ij})) + W_2(T(Fc_{ij} = fc_{ij}) \times W((Fc_{ij} = fc_{ij}))$$

Where $T(Fc_{ij} = fc_{ij})$ is the keyword context score which is a function of the matching percentage and is determined by the number of keyword matching for each token e.g. if two keywords out of four match this will be 50%. $W((Fc_{ij} = fc_{ij})$ are the weights associated with contexts that correspond to tokens of different keywords e.g. $fc_{1i}, fc_{2i}, fc_{3i}$ correspond to tokens of one, two and three keywords. $W_1$ and $W_2$ are weighting factors that can be used to control the contribution of threat score from keywords and keyword contexts.

In our experiments, we used one keyword, two keyword and three keyword tokens with a weight of 1, 2 and 3 respectively assigned to them. Feature space was divided into four groups with each group containing related features. Weights associated with contexts that corresponded to tokens and weighing factors ($W_1$ and $W_2$) were fine tuned to achieve higher accuracy.

### 3.4. Evaluation Measures

To evaluate the predictive accuracy of classifiers several measures have been proposed in literature [18]. The most simple measure is filtering accuracy namely percentage of messages classified correctly. More informative measures are recall and precision. Weighted error rate and weighted accuracy are a measure to evaluate filter accuracy. TCR is the relative cost of using the filter (and so having some false positives and some false negatives) to using no filter at all (and so having all the spam misclassified, but all the legitimate mail classified correctly). F-measure is the weighted harmonic mean of precision and recall. Since false positive are often more expensive than false negative, it is vital to compare the false positive rate of the classifier. The Receiver Operator Characteristics (ROC) curve is a graph to plot false positive against true positive, in which various threshold values are compared. We present, these measures in terms of threat and normal (spam and legitimate) e-mail as in table 1.

Let $n_{N \to N}$ be the number of normal messages classified as normal, $n_{N \to T}$ be the number of normal messages misclassified as threat, $n_{T \to T}$ be the number of threat messages classified as threat, $n_{T \to N}$ be the number of threat messages misclassified as normal and $\lambda$ be the weight on the accuracy of the classifier.

| Evaluation Measure | Evaluation Function |
|---|---|
| Accuracy | $Acc = \frac{n_{N \to N} + n_{T \to T}}{n_{N \to N} + n_{N \to T} + n_{T \to N} + n_{T \to T}}$ |
| Weighted Accuracy | $W_{Acc(\lambda)} = \frac{\lambda . n_{N \to N} + n_{T \to T}}{\lambda . (n_{N \to N} + n_{N \to T}) + n_{T \to N} + n_{T \to T}}$ |
| Error Rate | $Err_{Rate} = \frac{n_{N \to T} + n_{T \to N}}{n_{N \to N} + n_{N \to T} + n_{T \to N} + n_{T \to T}}$ |
| Weighted Error Rate | $W_{Err(\lambda)} = \frac{\lambda . n_{N \to T} + n_{T \to N}}{\lambda . (n_{N \to N} + n_{N \to T}) + n_{T \to N} + n_{T \to T}}$ |
| False Positive Rate | $FP_{Rate} = \frac{n_{N \to T}}{n_{N \to N} + n_{N \to T}}$ |
| False Negative Rate | $FN_{Rate} = \frac{n_{T \to N}}{n_{T \to N} + n_{T \to T}}$ |
| Recall | $r = \frac{n_{T \to T}}{n_{T \to N} + n_{T \to T}}$ |
| Precision | $p = \frac{n_{T \to T}}{n_{N \to T} + n_{T \to T}}$ |
| Total Cost Ratio | $Tc_{Ratio} = \frac{n_{T \to N} + n_{T \to T}}{\lambda . n_{N \to T} + n_{T \to N}}$ |
| ROC Curve | *True positive rate plotted against false positive rate* |

**Table** 1: Evaluation measures for Spam Filters

### 3.5. Results

Figure 2 plots the accuracy of Bayesian Single Keywords (BS), Bayesian weighted Multiple Keywords (BM) and Bayesian weighted Multiple Keywords with keyword Context matching (BMC) based filters as a function of data sizes. Accuracy of filters improved with increase in data size. Accuracy of BMC based filter improved by more than 4% while that of BM based filter improved by about 3%. Bayesian Single Keyword based filter did not improve much in accuracy with increase in training data size. Further, it was observed that the accuracy of BMC based filter improved greatly for higher percentage of spam.

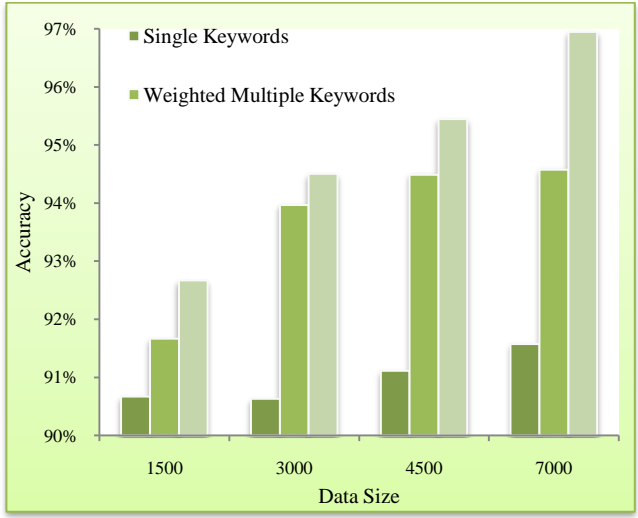

**Figure** 2: Accuracy vs. Data Size

In figure 3 accuracy of filters as a function of feature sizes is plotted. The accuracy of all filters improved with increase in the feature size of the classifier. For BMC based filters accuracy improved by 6% which is higher than that of filters based on BM and BS.

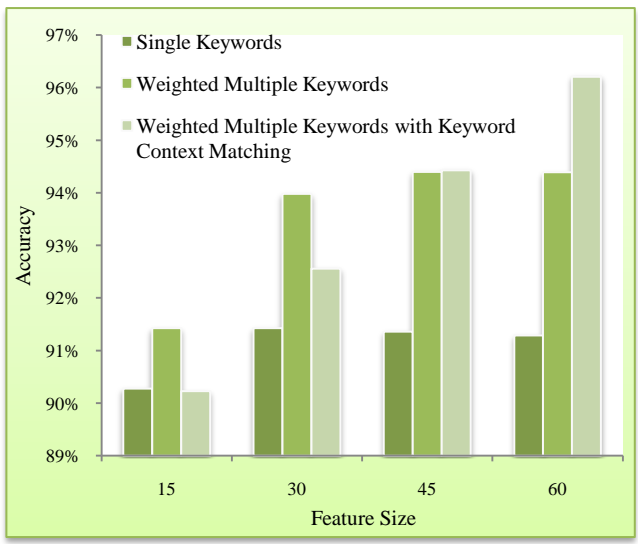

**Figure** 3: Accuracy vs. Number of Features

Performance of filters in terms of precision, recall, false positive rate, false negative rate, and F-measure is shown in table 2. This comparison is based on a data size of 7000 e-mails using 60 features for classification.

| *Measure* | *BS* | *BM* | *BMC* |
|---|---|---|---|
| *Precision ($p$)* | 90.76% | 97.56% | 98.10% |
| *Recall ($r$)* | 92.57% | 91.43% | 95.74% |
| *False Positive Rate ($FP_{Rate}$)* | 9.43% | 2.29% | 1.86% |
| *False Negative Rate ($FN_{Rate}$)* | 7.43% | 8.57% | 4.26% |
| *f-measure ($f1$)* | 91.65% | 94.40% | 96.91% |

**Table** 2: Performance in terms of Precision, Recall, FP, FN and f1

Appavu alias Balamurugan and his colleagues in their research [13] have reported the results of the use of several techniques in detecting threat e-mails from two different sets of e-mails (C1 and C2). These results are mentioned in table 3. Owing to the use of different data sets, it is not logical to compare the reported results with those obtained in our experiments. However, since in both, data set is created by farming threat e-mails and combining them with normal e-mails to form the e-mail corpus, comparison between the two can be drawn with a reasonable error.

| *Filtering Technique* | *Evaluation Measures in %age* | | | | | | | |
|---|---|---|---|---|---|---|---|---|
| | *A* | | *p* | | *r* | | *f1* | |
| | *C1* | *C2* | *C1* | *C2* | *C1* | *C2* | *C1* | *C2* |
| *ADI-IG* | 98.75 | 95.50 | 97.88 | 94.15 | 96.72 | 93.82 | 97.29 | 93.98 |
| *DT-IG* | 97.28 | 92.45 | 94.44 | 90.40 | 94.04 | 92.42 | 94.24 | 91.40 |
| *SVM-IG* | 95.65 | 92.10 | 97.78 | 94.05 | 80.89 | 87.40 | 88.54 | 90.60 |
| *NB-IG* | 94.93 | 86.35 | 91.64 | 93.83 | 82.20 | 74.20 | 86.66 | 82.87 |
| *ADI-TFV* | 99.20 | 99.10 | 97.10 | 96.75 | 96.72 | 96.45 | 96.91 | 96.59 |
| *DT-TFV* | 98.80 | 98.30 | 96.30 | 96.25 | 96.10 | 96.05 | 96.20 | 96.15 |
| *SVM-TFV* | 98.50 | 98.10 | 94.10 | 94.50 | 92.80 | 91.90 | 93.45 | 93.13 |
| *NB-TFV* | 94.41 | 89.35 | 90.54 | 96.20 | 79.89 | 78.40 | 84.88 | 86.39 |

**Table** 3: Performance in terms of Accuracy, Precision, Recall and f-measure

When comparing various evaluation measures of filters given in table 3 with those produced by our experiments given in table 2, it is clear that the performance of Naïve Bayesian filtering approaches used in our experiments are comparable and satisfactory. The precision of Bayesian weighted Multiple Keywords with keyword Context matching (BMC) based filter has remained higher than other filters.

In table 4, the weighted error rate obtained for various filters using different weight is compared.

| Weight ($\lambda$) | Weighted Error Rate | | |
|---|---|---|---|
| | BS | BM | BMC |
| 1 | 8.43% | 5.43% | 3.06% |
| 9 | 9.23% | 2.91% | 2.10% |
| 999 | 9.4% | 2.3% | 1.9% |

**Table** 4: Weighted Error rate with different values for $\lambda$

Although no Bayesian filter showed 100% predictive accuracy but the Bayesian weighted Multiple Keywords with keyword Context matching (BMC) based filter achieved nearly 97% accuracy. Highest false positive rate was shown by Bayesian Single Keywords (BS) based filter followed by Bayesian weighted Multiple Keywords (BM) and Bayesian weighted Multiple Keywords with keyword Context matching (BMC) based filter. The false positive rate of BMC based filter has remained as low as 1.86%. The f-measure of BMC based filter remained nearly 97%, highest of all others. Further, weighted error rate of BMC based filter is lowest in comparison to that of filters based on other approaches. This rate further reduced with higher values of weight.

Total Cost Ratio (TCR) based on data size of 7000 e-mails using 60 features with weights of 1, 9 and 999 for different filters is given in table 5. TCR being a unifying measure of spam recall and precision incorporates the cost difference between two error types namely false positive and false negative. Higher value of TCR indicates better performance of filter. Any value of TCR that is less than 1 indicates filters inefficiency and is cost wise worst than not using the filter at all. For $\lambda = 1$, in all three different Naïve Bayesian filter approaches TCR score remained more than 1. The Bayesian weighted Multiple Keywords with keyword Context matching (BMC) based filter scored highest TCR score followed by Bayesian weighted Multiple Keywords (BM) and Bayesian Single Keywords (BS) based filters. For $\lambda = 9$, again all filters achieved TCR greater than 1 but for $\lambda = 999$, TCR score for all filters dropped below 1. This is due to the very high penalty on normal mail misclassified as threat and none of the three filters managed to eliminate these errors completely. For a value of $\lambda = 999$ or its any other value for which TCR score drops below 1 despite higher values of precision and recall, it can be concluded that none it is better not to use the filter at all. But, depending upon the nature of the filter application like identification of threat e-mails a much higher value of $\lambda$ may not be desirable but high recall is desired even at some cost of precision.

| Weight ($\lambda$) | TCR Scores | | |
|---|---|---|---|
| | BS | BM | BMC |
| 1 | 5.932 | 9.210 | 16.355 |
| 9 | 1.083 | 3.431 | 4.768 |
| 999 | 0.011 | 0.043 | 0.0537 |

**Table** 5: TCR with different values for $\lambda$

In order to evaluate three different Naïve Bayesian filter approaches across the tradeoff between true positives and true negatives, we present the results through a discrete plot called the ROC graph shown in figure 4. The ROC curve is a graphical plot of spam accuracy or spam recall or true positive rate called formerly sensitivity on *Y-axis* against false positive rate or ham misclassification rate or the complement of ham recall formerly called 1-specificity on *X-axis*. Each point on the graph shown in figure 4 corresponds to an actual (*false positive rate, true positive rate*) pair achieved by particular Naïve Bayesian filter on a particular size of the data set. We have restricted the horizontal axis to a maximum of 0.2 i.e. 20% of the false positive rate and the vertical axis from 0.8 to improve readability. Points on the left side of the graph in the upper triangle and close to *X-axis* (for conservative classifiers) make positive classifications only with a strong evidence and thus make few false positive errors but often have low true positive rates as well. Such a point in figure 4 is point labeled as A. Points on the upper right side of the graph in the upper triangle (for liberal classifiers) make positive classifications with weak evidences and thus classify nearly all positives correctly but often have high false positive rates. Such a point in figure 4 is point labeled as B. It is difficult if not impossible some day to have a perfect classification i.e. a point at (0, 1). Any point near it is better than others. Such a point in figure 4 is labeled as C. The plotted results indicate that all used Naïve Bayesian filtering approach are characterized by points in the upper left quadrant in the graph but the Naïve Bayesian approach based on weighted Multiple Keywords with keyword

Context matching (BMC) is more to the northwest i.e. true positive rate is higher and false positive rate is lower.

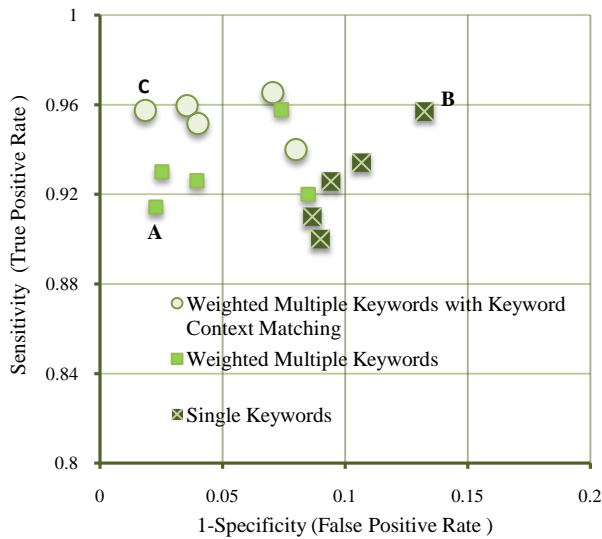

**Figure** 4: ROC curves of the filters

Overall, the Naïve Bayesian approach based on weighted Multiple Keywords with keyword Context matching (BMC) obtained the best results in our experiments both in spam and ham recall, however, the difference from other Naïve Bayesian approaches were small.

## 4. Conclusion

Naïve Bayesian filtering approaches namely single keywords, weighted multiple keywords and weighted multiple keywords with keyword context matching were evaluated in terms of various metrics on a threat e-mail corpus created by extracting data from sources that are very close to terrorism. None of these classification techniques showed 100% predicative accuracy but weighted multiple keywords with keyword context matching Naïve Bayesian filtering approach showed promising performance in terms of accuracy, ability to learn, and F-measure. Further, weighted error, false positive and false negative rates remained very low in comparison to that of filters based on other approaches. Since accuracy of filter evaluation is highly determined by relevance of training data, thus it is necessary to create a threat e-mail corpus to facilitate researchers in designing systems capable of automatically detecting and blocking terrorism threats by misuse of e-mail system. The openness of e-mail system and its use in unlawful activities makes filtering an active area for research and thus there exists a wide scope for development of new filters and improvements in the existing ones.

## Appendix

Some examples of the threat e-mail used in our corpus that can give readers an insight of the underlying data set, are hereunder.

### Example 1:

Today there will be bomb blast in parliament house and the US consulates in India at 11:46 am. Stop it if you could. Cut relations with the U.S.A. Long live Osama Finladen Asadullah Alkalfi.

### Example 2: (E-mail Intercepted by MessageLabs)

Attention Please
Make sure this gets to the manager as soon as possible, because this is the only way to pass this information to you and get this case settled. We have been paid to set an electronic explosive device (Bomb) in your hotel which we have done. But I feel like helping you people, I have a concrete evidence of this information on a tape record and the second tape contains the information and contact of our employer. I demand $130,000(USD) which must be paid before I could disclose any information to you. I need to settle my tem with this money so they can go back to their destinations. I traveled to Africa on a business trip but I have everything under my control. I will mail you the tapes but that will be after my boys have gone and am assured of your maximum co-operation.
Note: My employer has a secret agent working with you in your hotel. Therefore this information must not be known or exposed to anybody else my employer will sense betrayal and you know what that means. (I will not accept any apology if you people make any mistake).
Do will have a deal or not?
Reply this email as soon as possible
Mind you, time is essence.

### Example 3:

Dear Brother
We are fighting for your and our independence from the draconic occupation of <country name>. We need men, money and arms to continue with our struggle for freedom. You are welcome to fight along with us. If it is not possible for you to fight actively you can still be part of this freedom struggle by contributing money to <organization name>. We invite you to visit our website <web address>. Remember <country name> is our enemy and we have to fight for the independence of ourselves.

<name of person>
<organization name>

## Biographies

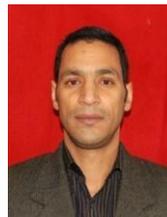

**M. Tariq Banday** was born in 1969. He did his M. Sc. and M. Phil. Degrees from the Department of Electronics, University of Kashmir, Srinagar, India in 1996 and 2008 respectively. He did advanced diploma course in computers and qualified UGC NET examination in 1997 and 1998. At present he is working as Assistant Professor in the Department of Electronics & Instrumentation Technology, University of Kashmir, Srinagar, India. He has to his credit several research publications in reputed journals and conference proceedings. He is a lifetime member of Computer Society of India and International Association of Engineers. His current research interests include Network Security, Internet Protocols and Network Architecture.

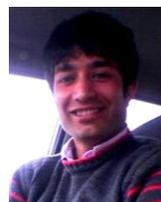

**Jameel A. Qadri** was born in 1972. He did his M.Sc. degree in Electronics and Post Graduate Diploma in Computer Applications from Department of Electronics, University of Kashmir, Srinagar, India in 1998 and 2000 respectively. He received M. Sc. Degree in Electronic Commerce from Middlesex University, Hendon Campus, London, UK in 2008. At present he is Lecturer in the Department of Information and Communication Technology in BC College of North West London, United Kingdom. His research articles have been published in journals and conference proceedings. His Current research interests are Internet Security, Knowledge Management and Web Accessibility.

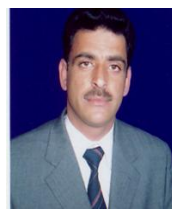

**Tariq Rashid Jan** was born in 1966. He did his M.Sc. and M. Phil. degrees in Statistics from Department of Statistics and Operations Research, Aligarh Muslim University, Aligarh, India in 1989 and 1994 respectively. He received his Ph. D. degree in Statistics from department of Statistics, University of Kashmir, Srinagar, India in 2004. At present he is Assistant Professor in the Department of Statistics, University of Kashmir, Srinagar. He has published work in international and national journals of repute. His Current research interests are bio-statistics, generalized distributions and quality control.

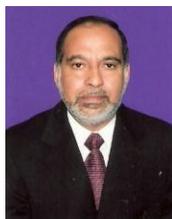 **Nisar A. Shah** was born in 1953. He did his M. Sc. and Ph. D. Degrees from the department of Physics, University of Kashmir, Srinagar, India in 1976 and 1981 respectively. At present he is working as Professor in the Department of Electronics & Instrumentation Technology, University of Kashmir. He has to his credit about 150 research publications which have been published in national and international journals of repute. He has supervised several research scholars in M. Phil. and Ph. D. programs. His current research interests include Digital Signal Processing and Network Security.